\documentclass[twocolumn,showpacs,preprintnumbers,amsmath,amssymb,prl,superscriptaddress]{revtex4-1}
\usepackage{bbm}
\usepackage{mathrsfs}
\usepackage{graphicx}
\usepackage{dcolumn}
\usepackage{bm}
\usepackage{amsmath}
\usepackage{amsfonts}
\usepackage{color}

\begin{document}

\title{Generation of Spin Currents by Magnetic Field in $\mathcal{T}$- and $\mathcal{P}$-Broken Materials}
\author{Jing Wang}
\affiliation{State Key Laboratory of Surface Physics and Department of Physics, Fudan University, Shanghai 200433, China}
\affiliation{Collaborative Innovation Center of Advanced Microstructures, Nanjing 210093, China}
\author{Biao Lian}
\author{Shou-Cheng Zhang}
\affiliation{Department of Physics, McCullough Building, Stanford University, Stanford, California 94305-4045, USA}

\begin{abstract}
Pure spin currents carry information in quantum spintronics and could play an essential role in the next generation low-energy-consumption electronics. Here we theoretically predict that the magnetic field can induce a quantum spin current without a concomitant charge current in metals without time reversal symmetry $\mathcal{T}$ and inversion symmetry $\mathcal{P}$ but respect the combined $\mathcal{PT}$ symmetry. It is governed by the magnetic moment of the Bloch states on the Fermi surface, and can be regarded as a spinful generalization of the gyrotropic magnetic effect in $\mathcal{P}$-broken metals. The effect is explicitly studied for a minimal model of an antiferromagnetic Dirac semimetal, where the experimental signature is proposed. We further propose candidate materials, including topological antiferromagnetic Dirac semimetals, Weyl semimetals, and tenary Heusler compounds.
\end{abstract}

\date{\today}


\maketitle

\paragraph{Introduction.}

Pure spin currents, which consists of opposite spins moving with opposite velocities, are invariant under the time reversal. They carry information via spins instead of charges and play an important role in modern quantum spintronics~\cite{zutic2004,macdonald2011}. Due to their intrinsic nondissipative nature, tremendous efforts have been made to generate and manipuate pure spin currents in semiconductors. A pure spin current can be induced by an electric field via spin Hall effect~\cite{hirsch1999,murakami2003,sinova2015}, or by optical injection~\cite{bhat2000,stevens2003,zhao2006} in semiconductors, which rely on the spin-orbit coupling and inversion asymmetry of the system. Recently, several novel magnetotransport phenomena have been predicted in topological Weyl semimetals, such as the chiral magnetic effect~\cite{son2012,zyuzin2012,vazifeh2013,chang2015a,chang2015b,zhong2016,ma2015,goswami2015}, where in the presence of a static magnetic field $\mathbf{B}$, an electric charge current $\mathbf{J}\parallel\mathbf{B}$ can be driven by an electric field $\mathbf{E}\parallel\mathbf{B}$. This motivates us to ask whether a magnetic field $\mathbf{B}$ alone could drive a pure spin current in certain systems such as topological materials, which would be important both for fundamental and technological interest. The positive answer to this question may lead to electronic and spintronic applications for topological materials.

In this paper we present a new effect of pure spin current induced by solely a magnetic field $\mathbf{B}$. We will demonstrate that metals without time-reversal symmetry $\mathcal{T}$ and inversion symmetry $\mathcal{P}$ but respect the combined $\mathcal{PT}$ symmetry can have such a quantum spin current induced by an oscillating $\mathbf{B}$ field without a concomitant charge current, which is determined by the intrinsic magnetic moment of the Bloch states on the Fermi surface. The effect is explicitly studied for a minimal model of an antiferromagnetic Dirac semimetal, where the experimental implication is proposed. The spin current flowing direction is controlled by the external $\mathbf{B}$ field. This is in sharp contrast to the spin Hall effect where a pure Hall spin current is induced by a electric field, which is due to the geometric Berry curvature~\cite{sinova2015}. We further discuss the optimal conditions for observing this effect and propose candidate three-dimensional materials.

Our study builds upon seminal work by Zhong \emph{et al.}~\cite{zhong2016} and Ma \emph{et al.}~\cite{ma2015}, which predicted a charge current induced by an external magnetic field in $\mathcal{P}$-broken metals, named gyrotropic magnetic effect (GME). It can be regarded as a low-frequency limit of the natural gyrotropy~\cite{landau1984}. The effect of a pure spin current generated by $\mathbf{B}$ field presented here can be regarded as a spinful generalization of GME, therefore dubbed as \emph{spin GME}.

\paragraph{General theory.}

We begin with the general symmetry analysis of spin GME by considering the linear response of spin current $\boldsymbol{\mathcal{J}}$ induced by $\mathbf{B}$,
\begin{equation}\label{spincurrent}
\mathcal{J}_{ij} = \gamma_{ijl}B_l.
\end{equation}
Here $i$ denotes the spin directions, $j,l$ denote the spatial directions, and $i,j,l=1,2,3$. $\boldsymbol{\mathcal{J}}$ is a rank-$2$ pseudotensor, which is even under $\mathcal{T}$ and odd under $\mathcal{P}$; while $\mathbf{B}$ is $\mathcal{T}$ odd and $\mathcal{P}$ even. Therefore according to Eq.~(\ref{spincurrent}), the rank-$3$ tensor coefficient $\gamma_{ijl}$ can be nonzero only if both $\mathcal{T}$ and $\mathcal{P}$ symmetries are broken.
Similarly, GME can be discussed by posting the linear relation between charge current $\mathbf{J}$ and $\mathbf{B}$~\cite{zhong2016,ma2015},
\begin{equation}
J_i=\alpha_{ij}B_j.
\end{equation}
Both $\mathbf{J}$ and $\mathbf{B}$ are odd under $\mathcal{T}$, while $\mathbf{J}$ is $\mathcal{P}$ odd and $\mathbf{B}$ is $\mathcal{P}$ even, so the rank-$2$ tensor $\alpha_{ij}$ is nonzero only if both $\mathcal{P}$ and $\mathcal{PT}$ symmetries are broken.
The explicit form of both ${\gamma}$ and ${\alpha}$ can be obtained by the Kubo formula in the \emph{uniform} limit (i.e., $\mathbf{q}=0$ before $\omega\rightarrow0$), and both of them are material properties.

To have an intuitive picture of these effects, here we give a unified derivation of both spin GME and GME. Since we are considering the regime where $\hbar\omega\ll\epsilon_{\text{gap}}$, it is legitimate to use the Boltzmann equation. The combined $\mathcal{PT}$ symmetry reverses spin and orbital angular momentum, and ensures that all of the energy bands are two-fold degenerate. Generically, we consider a doubly degenerate band with energy dispersion $\epsilon_{\mathbf{k}}$ and basis wavefunctions $(|u_{\mathbf{k}}^+\rangle,|u_{\mathbf{k}}^-\rangle)^T$ in the momentum $\mathbf{k}$-space. $+/-$ denotes the two degenerate bands. As is guaranteed by the $\mathcal{PT}$ symmetry, the electron spin operator projected into these two basis is generically a vector $2\times 2$ matrix $\mathbf{s}_\mathbf{k}=\sum_i\mathbf{s}^i_{\mathbf{k}}\sigma_i, (i=x,y,z)$, where $\mathbf{s}^i_\mathbf{k}$ is a vector field in $\mathbf{k}$-space, and $\sigma_i$ is the $i$-th Pauli matrix. The reduced $1$-particle density matrix $n_{\mathbf{k}}$ is also a $2\times2$ matrix, which reduces to the Fermi-Dirac distribution when the system is in equilibrium.
The spin current density is given by the integral of the spin $\mathbf{s}_{\mathbf{k}}$ and physical velocity $\mathbf{v}_{\mathbf{k}}$ of the electrons, weighed by $n_{\mathbf{k}}$:
\begin{equation}
\boldsymbol{\mathcal{J}}=-e\int_\mathbf{k} \mathrm{Tr}\left(n_{\mathbf{k}}\mathbf{s}_{\mathbf{k}}\right)\mathbf{v}_{\mathbf{k}}.
\end{equation}
Here we have avoided the subtlety in the definition of a spin current~\cite{shi2006}. $\int_{\mathbf{k}}\equiv\int d^3\mathbf{k}/(2\pi)^3$ and the integral is over the Brillouin zone (BZ). $-e$ is the electron charge, $\mathbf{v}_{\mathbf{k}}=\hbar^{-1}\nabla_\mathbf{k}\epsilon_{\mathbf{k}}$ is the group velocity of the band electron. In general, via a local basis rotation we can make $\mathbf{s}^x_\mathbf{k}=\mathbf{s}^y_\mathbf{k}=0$ everywhere in the $\mathbf{k}$-space, i.e., the spin operator takes the form $\mathbf{s}_\mathbf{k}=\mathbf{s}^z_\mathbf{k}\sigma_z$.
In contrast, the charge current density is given by
\begin{equation}
\mathbf{J}=-e\int_\mathbf{k} \mathrm{Tr}\left(n_{\mathbf{k}}\right)\mathbf{v}_{\mathbf{k}}.
\end{equation}

In the presence of an external magnetic field, the band electron responses via its magnetic moment $\mathbf{m}_\mathbf{k}$, which is a $2\times2$ matrix. Semiclassically, it can be expressed as $\mathbf{m}_\mathbf{k}=-(eg_s/2m_e)\mathbf{s}_\mathbf{k}+\boldsymbol{\ell}_{\mathbf{k}}$, where $g_s$ is the electron spin $g$-factor, and $m_e$ is the electron mass. $\boldsymbol{\ell}_{\mathbf{k}}$ is the orbital magnetic moment and has the matrix elements $\boldsymbol{\ell}_{\mathbf{k}}^{\mu\varrho}=(e/2)\mbox{Im}\langle\nabla_\mathbf{k}u_{\mathbf{k}}^{\mu}|\times(H_\mathbf{k}-\epsilon_{\mathbf{k}}) |\nabla_\mathbf{k}u_{\mathbf{k}}^{\varrho}\rangle$ for $\mu,\varrho=\pm$. For simplicity, here we neglect the off-diagonal part of $\boldsymbol{\ell}_{\mathbf{k}}$ and assume it takes the form as $\boldsymbol{\ell}_{\mathbf{k}}=\boldsymbol{\ell}^{z}_{\mathbf{k}}\sigma_z$, and hence the total magnetic moment becomes $\mathbf{m}_\mathbf{k}=[-(eg_s/2m_e)\mathbf{s}^z_\mathbf{k}+\boldsymbol{\ell}^z_{\mathbf{k}}]\sigma_z
=\mathbf{m}^z_\mathbf{k}\sigma_z$. The band energy has a correction from the magnetic moment as $\tilde{\epsilon}_{\mathbf{k}\pm}=\epsilon_\mathbf{k}\mp\mathbf{m}^z_\mathbf{k}\cdot\mathbf{B}$. Therefore $n_{\mathbf{k}}$ can be approximated as diagonal, $n_\mathbf{k}=\mbox{diag}(n_{\mathbf{k}+},n_{\mathbf{k}-})$. Now in the absence of an external electric field, the semiclassical equations of motion of the Bloch electron are~\cite{xiao2010}
\begin{equation}\label{motion}
\begin{aligned}
\dot{\mathbf{r}}_{\pm} &= \nabla_{\mathbf{k}}\tilde{\epsilon}_{\mathbf{k}\pm}-\dot{\mathbf{k}}_{\pm}\times\boldsymbol{\Omega}_{\mathbf{k}\pm},
\\
\dot{\mathbf{k}}_{\pm} &= -\nabla_{\mathbf{r}}\tilde{\epsilon}_{\mathbf{k}\pm}-e\dot{\mathbf{r}}_{\pm}\times\mathbf{B}.
\end{aligned}
\end{equation}
Here $\boldsymbol{\Omega}_{\mathbf{k}\pm}=-\text{Im}\langle\nabla_\mathbf{k}u_{\mathbf{k}}^\pm|\times |\nabla_\mathbf{k}u_{\mathbf{k}}^\pm\rangle$ are the Berry curvatures of the two degenerate bands, and the $\mathcal{PT}$ symmetry ensures $\boldsymbol{\Omega}_{\mathbf{k}+}=-\boldsymbol{\Omega}_{\mathbf{k}-}$. In the relaxation time approximation the Boltzmann equation for the distribution of electrons is
\begin{equation}
\partial_t n_{\mathbf{k}\pm}+\dot{\mathbf{r}}_{\pm}\nabla_{\mathbf{r}} n_{\mathbf{k}\pm}
+\dot{\mathbf{k}}_{\pm}\nabla_{\mathbf{k}} n_{\mathbf{k}\pm}=\frac{f^0_{\mathbf{k}\pm}-n_{\mathbf{k}\pm}}{\tau},
\end{equation}
where $\tau$ is the relaxation time, and $f^0_{\mathbf{k}\pm}=f(\tilde{\epsilon}_{\mathbf{k}})\simeq f(\epsilon_{\mathbf{k}})\mp\mathbf{m}^z_{\mathbf{k}}\cdot\mathbf{B}(\partial f/\partial\epsilon_{\mathbf{k}})$ is the instantaneous equilibrium distribution. Although this approximation does not take into account the self-energy effects, it turns out to be qualitatively correct and quantitatively quite accurate. $f(\epsilon_{\mathbf{k}})$ is the Fermi-Dirac distribution for the band energy $\epsilon_{\mathbf{k}}$. Consider the driving magnetic field $\mathbf{B}(t)=\mathbf{B}_0 e^{-i\omega t}$ which is uniform in space but oscillates harmonically in time with frequency $\omega$. Since we are interested in computing the response to linear order in $\mathbf{B}$, Eq.~(\ref{motion}) yields $\dot{\mathbf{k}}_\pm=-e\mathbf{v}_\mathbf{k}\times\mathbf{B}$.
Therefore the Boltzmann equation to the linear order of $\mathbf{B}$ reduces to
\begin{equation}
f^1_{\mathbf{k}\pm}=\frac{\pm i\omega\tau}{i\omega\tau-1}\frac{\partial f}{\partial\epsilon_{\mathbf{k}}}\mathbf{m}^z_\mathbf{k}\cdot\mathbf{B},
\end{equation}
where $f^1_{\mathbf{k}\pm}=n_{\mathbf{k}\pm}-f^0_{\mathbf{k}\pm}$. It is noted that if $\omega\rightarrow0$, then $f^1_{\mathbf{k}\pm}\rightarrow0$, which means a static $\mathbf{B}$ field only modifies the equilibrium state, therefore both $\boldsymbol{\mathcal{J}}$ and $\mathbf{J}$ associated with $f^0_{\mathbf{k}}$ vanish. The pure spin current induced by the oscillating $\mathbf{B}$ field is
\begin{equation}
\boldsymbol{\mathcal{J}}=\frac{2ie\omega\tau}{1-i\omega\tau}
\int_{\mathbf{k}}\frac{\partial f}{\partial\epsilon_{\mathbf{k}}}(\mathbf{m}^z_{\mathbf{k}}\cdot\mathbf{B})\mathbf{s}^z_\mathbf{k}\mathbf{v_k}.
\end{equation}
The magnetic field $\mathbf{B}=\nabla\times\mathbf{A}$ is linear in $\mathbf{q}$ for an optical field, therefore $\mathcal{J}\propto qA$, which is similar to the linear optical effects of pure spin currents~\cite{wang2008,wang2012b}. In general, with conserved $\mathcal{PT}$ symmetry, $\mathbf{m}_\mathbf{k}$ is a generic $2\times2$ traceless Hermitian  matrix, and in this case, the spin current formula becomes
\begin{equation}
\boldsymbol{\mathcal{J}}=\frac{i\omega\tau e}{1-i\omega\tau}\int_{\mathbf{k}}\frac{\partial f}{\partial\epsilon_\mathbf{k}}\mbox{Tr}\left(
\mathbf{s}_\mathbf{k}\mathbf{m}_\mathbf{k}\right)\cdot\mathbf{B}\mathbf{v}_\mathbf{k}.
\end{equation}
This is the key result of this paper~\cite{note}. At zero temperature, $\partial f/\partial\epsilon_{\mathbf{k}}=-\delta^3(\mathbf{k}-\mathbf{k_F})/\hbar|\mathbf{v_k}|$, we obtain the Fermi surface formula
\begin{equation}\label{spin_GME}
\gamma_{ijl}=\frac{2i\omega\tau}{i\omega\tau-1}\frac{e}{(2\pi)^2h}\sum_{a}\oint_{S_a}\sum_{n=1}^3
{s}^n_{\mathbf{k},i}{m}^n_{\mathbf{k},l}\hat{v}_{F,j}.
\end{equation}
Here $S_a$ is the $a$-th Fermi surface sheet in band, $\hat{\mathbf{v}}_F$ is the Fermi surface normal at $\mathbf{k}_F$, $h$ is the Plank constant, and ${s}^n_{\mathbf{k},i},{m}^n_{\mathbf{k},i},\hat{v}_{F,i}$ denotes the $i-$th component of $\mathbf{s}^n_{\mathbf{k}},\mathbf{m}^n_{\mathbf{k}},\mathbf{\hat{v}}_{F}$, respectively. With $\mathcal{P}$ symmetry present, one has $\mathbf{m}_{-\mathbf{k}}=\mathbf{m_k}$, $\mathbf{s}_{-\mathbf{k}}=\mathbf{s_k}$ and $\hat{\mathbf{v}}_{F}(-\mathbf{k}_F)=-\hat{\mathbf{v}}_{F}(\mathbf{k}_F)$, leading to $\gamma_{ijl}=0$. While with $\mathcal{T}$ symmetry present, $\mathrm{Tr}(\mathbf{s}_{-\mathbf{k}}\mathbf{m}_{-\mathbf{k}})=\mathrm{Tr}(
\mathbf{s}_\mathbf{k}\mathbf{m}_\mathbf{k})$ and $\hat{\mathbf{v}}_{F}(-\mathbf{k}_F)=-\hat{\mathbf{v}}_{F}(\mathbf{k}_F)$, also leading to $\gamma_{ijl}=0$. Therefore, the spin GME can only occur if both $\mathcal{T}$ and $\mathcal{P}$ are broken. Moreover, the spin GME is determined by the intrinsic magnetic moment of the Bloch states on the Fermi surface, therefore it is in general non-quantized and vanishes trivially for insulators.

In contrast, the charge current $\mathbf{J}$ vanishes due to $\mathrm{Tr}(n_{\mathbf{k}})\propto\mathrm{Tr}(\mathbf{m_k})=0$ in the presence of $\mathcal{PT}$ symmetry. Therefore with broken $\mathcal{PT}$ symmetry, the energy bands are non-degenerate and $\mathrm{Tr}(\mathbf{m_k})\neq0$ in general. The charge current is nonzero and has the form as $\mathbf{J}\propto\int_{\mathbf{k}}(\partial f/\partial\epsilon_{\mathbf{k}})\mathbf{v_k}\mathrm{Tr}(\mathbf{m}_{\mathbf{k}})
\cdot\mathbf{B}$.

As discussed in Ref.~\cite{zhong2016,ma2015}, the GME is the low frequency limit of natural gyrotropy in $\mathcal{P}$-broken metals. In fact, the spin GME discussed above can be viewed as the low frequency limit of optical injection of ballistic currents~\cite{zhao2006}. The key element there is for breaking the $\mathbf{k}$-space symmetry of the optical excitation. With broken $\mathcal{T}$ and $\mathcal{P}$ but conserved $\mathcal{PT}$, $\epsilon_{\mathbf{k},\uparrow}=\epsilon_{\mathbf{k},\downarrow}$ but $\epsilon_{\mathbf{k},\uparrow}\neq\epsilon_{-\mathbf{k},\uparrow}$. Therefore the transition amplitude $A(\mathbf{k})\neq A(-\mathbf{k})$, which results in the transition probability $P(\mathbf{k})\equiv |A(\mathbf{k})|^2\neq P(-\mathbf{k})$ in most cases.

\paragraph{Minimal model.}

\begin{figure}[t]
\begin{center}
\includegraphics[width=2.8in]{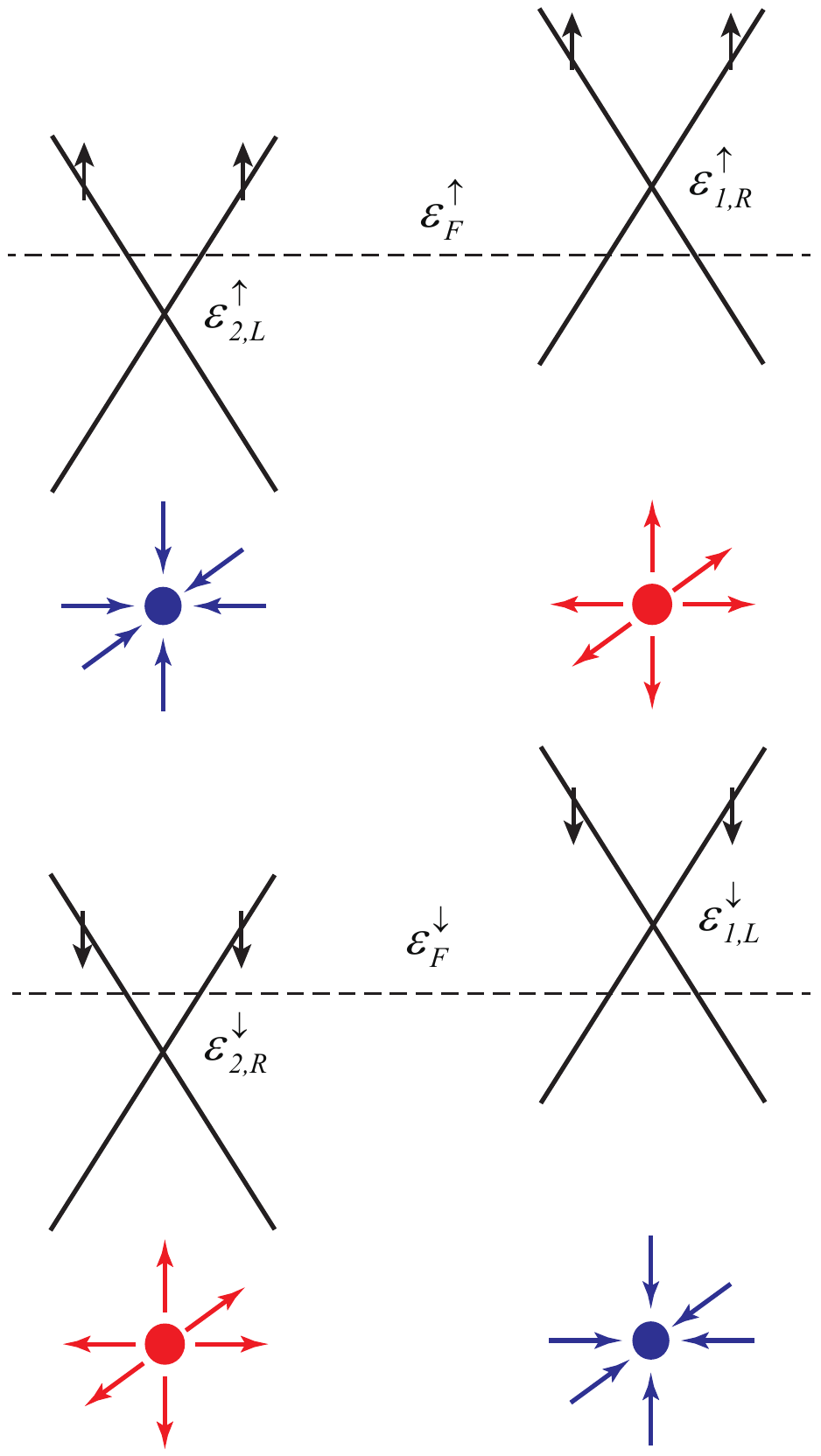}
\end{center}
\caption{(Color online) Pure spin currents generated by magnetic field in a $\mathcal{P}$, $\mathcal{T}$-broken but $\mathcal{PT}$-invariant Dirac semimetal with minimal two DPs labeled as $1,2$. The spin is independent of $\mathbf{k}$, and each Dirac node is decoupled into two Weyl nodes with opposite spins, chiralities and monopole charges of Berry curvature. The spin and chirality is denoted as spin up ($\uparrow$) and spin down ($\downarrow$), $L$ and $R$. Two Dirac nodes are denoted as four Weyl nodes, with the schematics of Berry monopole and antimonopole. Both $\mathcal{P}$ and $\mathcal{T}$ symmetry are now broken, leading to the different energy of the Dirac nodes $\varepsilon_1\neq\varepsilon_2$. The combined $\mathcal{PT}$ symmetry is conserved, leading to $\varepsilon_{1,L}^{\uparrow}=\varepsilon_{1,R}^{\downarrow}$ and $\varepsilon_{2,R}^{\uparrow}=\varepsilon_{2,L}^{\downarrow}$. The Fermi pockets are equilibrated with a common Fermi level, and an oscillating $\mathbf{B}$-field drives the spin current.}
\label{fig1}
\end{figure}

Now we consider the spin GME for a concrete model of $\mathcal{PT}$-invariant metals. The simple system adopted here is the minimal Dirac semimetal~\cite{young2012,wang2012,wang2013} with only two Dirac points (DPs) as shown in Fig.~\ref{fig1}, where there are only four bands are close to $\epsilon_F$ and the couplings to more distant bands can be neglected. With broken both $\mathcal{P}$ and $\mathcal{T}$, the two DPs are at different energies, but we assume $\epsilon_F$ is close to both. Now the Fermi surface consists of two pockets surrounding isotropic DPs. Around each DPs the effective Hamiltonian is
\begin{equation}
\mathcal{H}_{\text{Dirac}}=\varepsilon_{\nu}+\eta_{\nu}v_F\left(k_x\tau_x+k_y\tau_y+k_z\tau_z\sigma_z\right),
\end{equation}
where $\nu=1,2$ labels the DP, $\varepsilon_{\nu}$ is its energy, $\eta_{1}=1$, $\eta_2=-1$, $v_F$ is the Fermi velocity, $\mathbf{k}$ is expanded from the DP, $\tau_{x,y,z}$ and $\sigma_z$ are Pauli matrices for orbital and spin basis, respectively. Each Dirac cone can be decoupled into two Weyl cones with opposite chiralities and spins, and the total chirality vanishes. Take $\eta_1=1$ for example, $\mathcal{H}_{1,\text{Weyl}}^{\pm}=\varepsilon_1+v_F\left(k_x\tau_x+k_y\tau_y\pm k_z\tau_z\right)$. As shown in Fig.~\ref{fig1}, the two Weyl points have opposite chiralities and can be labeled as left- and right-handed Weyl nodes in the subscripts. The eigenvalues are $\epsilon_{1,R}=\varepsilon_1\pm v_F|\mathbf{k}|$, and $\epsilon_{1,L}=\varepsilon_1\pm v_F|\mathbf{k}|$. The opposite Weyl nodes at each DP has opposite magnetic moment, therefore the magnetic moment is traceless and off-diagonal term vanishes. In fact, the magnetic moment (we focus here on the orbital contribution for simplicity) for each DP is calculated as $\boldsymbol{\ell}_{\mathbf{k}\nu}=-\eta_{\nu}(ev_F/2k)\hat{\mathbf{k}}\sigma_z$. The spin $\mathbf{s_k}=\mathbf{s}\sigma_z$ is independent of $\mathbf{k}$. Take $\mathbf{s}=s\hat{z}$ for example,
only the $z$-component trace piece $\bar{\gamma}^z\delta_{jl}$ survives in Eq.~(\ref{spin_GME}); in the $\omega\tau\gg1$ limit each Dirac pocket contributes
\begin{equation}
\bar{\gamma}^z_\nu=\pm\frac{2}{3}\frac{e^2}{h^2}\eta_\nu v_Fk_F=\frac{2}{3}\frac{e^2}{h^2}\eta_\nu(\varepsilon_\nu-\varepsilon_F),
\end{equation}
where the plus (minus) sign depends on $\varepsilon_\nu>\varepsilon_F$ ($\varepsilon_\nu<\varepsilon_F$). The spin GME coefficient is obtained by summing over $\nu$ as $\bar{\gamma}^z=(2e^2/3h^2)\sum_\nu\eta_\nu\varepsilon_\nu$. The spin current induced by the magnetic field is,
\begin{equation}\label{sc_dirac}
\boldsymbol{\mathcal{J}}=(2e^2/3h^2)(\varepsilon_1-\varepsilon_2)\mathbf{s}\mathbf{B}.
\end{equation}

The spin current discussed in Eq.~(\ref{sc_dirac}) is related to the charge current for different spin components. The GME is absent due to the $\mathcal{PT}$ symmetry. However, if we further consider the GME for different spin components, we find that for Weyl nodes with spin up ($|\uparrow\rangle$) as shown in top row of Fig.~\ref{fig1}, these two Weyl nodes have opposite chiralities and opposite orbital moments, and the GME charge current is $\mathbf{J}^{\uparrow}=(e^2/3h^2)(\varepsilon_{1,R}^{\uparrow}-\varepsilon_{2,L}^{\uparrow})\mathbf{B}$. Similarly, the GME current for spin down $|\downarrow\rangle$ is $\mathbf{J}^{\downarrow}=(e^2/3h^2)(\varepsilon_{2,R}^{\downarrow}-\varepsilon_{1,L}
^{\downarrow})\mathbf{B}$~\cite{zhong2016}. The conserved $\mathcal{PT}$ symmetry leads to $\varepsilon_{2,R}^{\downarrow}=\varepsilon_{2,L}^{\uparrow}$ and $\varepsilon_{1,R}^{\downarrow}=\varepsilon_{1,L}^{\uparrow}$. Therefore the GME charge current $\mathbf{J}=\mathbf{J}^{\uparrow}+\mathbf{J}^{\downarrow}=0$, while the pure spin current $\boldsymbol{\mathcal{J}}_3=\mathbf{J}^{\uparrow}-\mathbf{J}^{\downarrow}\neq0$. We emphasize that conserving $\mathcal{PT}$ is not a necessity for spin GME, only broken both $\mathcal{P}$ and $\mathcal{T}$ is required. In $\mathcal{P}$-broken (polar or chiral) metals, GME current is present. With further $\mathcal{T}$ breaking, the charge current component for opposite spin is not equal to each other, leading to polarized spin currents.

The spin current is generated by the oscillating $\mathbf{B}$ field, however, it is suppressed by scattering when $\omega\ll1/\tau$, and becomes strong when $\omega\gg1/\tau$. Besides, Eq.~(\ref{sc_dirac}) indicates the spin current flowing direction is along $\mathbf{B}$ field, therefore, one can generate the longitudinal component ($\mathbf{s}\parallel\mathbf{B}$) and transverse component ($\mathbf{s}\perp\mathbf{B}$) separately. This is due to the isotropic Dirac nodes with spherical Fermi surface. In general, the $\gamma$ tensor has 27 independent terms, leading to generation of more complicated form of spin currents. However, the crystal point symmetry will set many terms to be zero or nonindependent. $\gamma$ is a rank-$3$ tensor, hence crystal symmetries impose constraints of the
form $\gamma_{ijl}=\mathcal{R}_{i}^{\ i'}\mathcal{R}_{j}^{\ j'}\mathcal{R}_{l}^{\ l'}\gamma_{i'j'l'}$, where $\mathcal{R}$ is an orthogonal matrix describing the point group.

\paragraph{Materials.} The magnetic moment is approximately proportional to the Berry curvature of the Bloch states and is often concentrated in a few regions of $\mathbf{k}$-space where two or more bands get close in energies~\cite{yao2008}. Therefore, the Dirac and Weyl materials with broken $\mathcal{T}$ and $\mathcal{P}$ are ideal candidates for observing the spin GME predicted in this work. We propose three classes of candidate materials: antiferromagnetic Dirac semimetals, magnetic and noncentrosymmetric Weyl semimetals, and half-Heusler compounds.

The first interesting candidate is the recently predicted Dirac semimetals in the antiferromagnetic CuMnAs material class~\cite{tang2016,wadley2016}. The antiferromagnetic order on Mn atoms breaks both $\mathcal{T}$ and $\mathcal{P}$ whereas $\mathcal{PT}$ still holds. It hosts four massless Dirac fermions, which is grouped into two pairs with different energies of DPs. One pair of the DPs is located along the high-symmetric X-U line and is protected by nonsymmorphic symmetry, while the other pair is in the interior of the BZ. The effective model of this system is just a two copies of the minimal model discussed above. The calculated splitting $|\varepsilon_1-\varepsilon_2|\sim0.01$~eV. For an estimation, take $B_0=3$~G, we get the magnitude of the spin current density is $18.6$~nA/$\mu$m$^2$ per Dirac node pair, in the range accessible by transport or optical experiments. The spin currents generated here can be detected by the conventional Faraday/Kerr rotation due to spin polarized electrons accumulated at the sample edges~\cite{kato2004,stern2008}, or by electric measurement via inverse spin Hall effect~\cite{sinova2015} to convert into voltage signal~\cite{valenzuela2006}, or by second harmonic generation~\cite{wang2010,weraka2010}. The optical frequency in the measurements should be in the range from infrared to $1/\tau$, which depends on the quality of materials. Although the relaxation time in CuMnAs has not been determined to the best of our knowledge, the experiments for an analogous Dirac semimetal material Cd$_3$As$_2$ shows that $\tau\sim2\times10^{-10}$~s~\cite{liang2015}, which indicates the observation of the predicted spin current via Faraday rotation is promising.

Another candidate materials are the magnetic and noncentrosymmetric Weyl semimetals predicted in the \emph{R}AlGe family of compounds~\cite{chang2016}, where \emph{R} denotes rare-earth element. These materials breaks $\mathcal{P}$, and $\mathcal{T}$ with ferromagnetic order (\emph{R} = Ce, Pr) or antiferromagnetic order (\emph{R} = Sm, Gd). Take PrAlGe for example, the magnetization breaks the mirror symmetry, which misaligns the Weyl points with opposite chiralities. The energy difference between the inequivalent Weyl points is about $0.02$~eV, which will give rise to large enough spin current density in experiments.

Last not the least, the large family of ternary half-Heusler compounds is a prime
candidate for multifunctional topological states, which combines topological and symmetry-breaking orders~\cite{chadov2010}. The crystal structure of these materials is non-centrosymmtric. For example, coexistence of bulk magnetism~\cite{canfield1991} and superconductivity are found in topological half-Heusler semimetals \emph{R}PtBi (\emph{R} = Y, Sm, Gd, Tb, Dy, Ho, Er, Tm, and Lu)~\cite{nakajima2015}. The broken $\mathcal{T}$ and $\mathcal{P}$ in these materials may realize the magnetic field induced spin currents.

In summary, we predict the magnetic field induced spin currents in antiferromagnetic Dirac semimetals, which is expected to have a great impact for electronic and spintronic applications of topological materials, particularly in view of recent developments in antiferromagnetic spintronics~\cite{macdonald2011,jungwirth2016}. Such a prediction is generic for topological materials with broken $\mathcal{T}$ and $\mathcal{P}$ symmetry. The general form of $\gamma$ tensor coefficient is determined by the point group symmetry~\cite{tensor}, and will be studied in future work. With efficient \emph{ab initio} calculations of magnetic moments~\cite{lopez2012}, we expect the predicted effect here in more $\mathcal{T}$- and $\mathcal{P}$-broken materials could be identified.

\begin{acknowledgments}
\emph{Acknowledgments.} We are grateful to Q. Zhou and P. Yu for valuable discussions. This work is supported by the National Thousand-Young-Talents Program and in part by Fudan University Initiative Scientific Research Program. S.C.Z. is supported by the US Department of Energy, Office of Basic Energy Sciences, Division of Materials Sciences and Engineering, under Contract No.~DE-AC02-76SF00515 and in part by the NSF under grant No.~DMR-1305677.
\end{acknowledgments}


\begin{thebibliography}{41}
\bibitem [{not()}]{note}%
  \BibitemOpen
  \href@noop {} {}\bibinfo {note} {The spin precession due to the magnetic
  field can be neglected, as long as precession frequency
  $\omega_0\ll\omega\ll\epsilon_{\text{gap}}$. For an estimation, take
  $B_0\sim3$~G, $\omega_0\sim10^7/\text{s}$. Now with $\tau\sim10^{-10}$~s, to
  observe the spin GME, the frequency of $\mathbf{B}$ field
  $\omega\gtrsim1/\tau\gg\omega_0$, validates the above condition.}\BibitemShut
  {Stop}%
\bibitem [{ten()}]{tensor}%
  \BibitemOpen
  \href@noop {} {}\bibinfo {note} {There is no isotropic tensor of
  rank-3.}\BibitemShut {Stop}%
\end{thebibliography}
\end{document}